\documentclass[%
 reprint,
 superscriptaddress,
 amsmath,amssymb,
 aps,
 prb,
]{revtex4-2}

\usepackage{graphicx}
\usepackage{dcolumn}
\usepackage{bm}
\usepackage[colorlinks=true,allcolors=blue]{hyperref}%

\begin{document}

\title{Photocarrier Transport of Ferroelectric Photovoltaic Thin Films Detected by the Magnetic Dynamics of Adjacent Ferromagnetic Layers}

\author{Yujun Zhang}
\email{zhangyujun@ihep.ac.cn}
\thanks{These authors contributed equally to this work.}
\affiliation{Institute of High Energy Physics, Chinese Academy of Sciences, Yuquan Road 19B, Shijingshan District, Beijing, 100049, China}
\affiliation{Department of Material Science, Graduate School of Science, University of Hyogo, Ako, Hyogo 678-1297, Japan}
\author{Ji Ma}
\thanks{These authors contributed equally to this work.}
\affiliation{Kunming University of Science and Technology, Kunming, Yunnan, China}
\affiliation{State Key Lab of New Ceramics and Fine Processing, School of Materials Science and Engineering, Tsinghua University, Beijing 100084, China}
\author{Keisuke Ikeda}
\affiliation{Institute for Solid State Physics, University of Tokyo, 5-1-5 Kashiwanoha, Chiba 277-8581, Japan}
\author{Yasuyuki Hirata}
\affiliation{Institute for Solid State Physics, University of Tokyo, 5-1-5 Kashiwanoha, Chiba 277-8581, Japan}
\author{Kohei Yamagami}
\affiliation{Institute for Solid State Physics, University of Tokyo, 5-1-5 Kashiwanoha, Chiba 277-8581, Japan}
\author{Christian Sch\"u\ss ler-Langeheine}
\affiliation{Helmholtz-Zentrum Berlin f\"ur Materialien und Energie GmbH, Albert-Einstein-Stra\ss e 15, 12489 Berlin, Germany}
\author{Niko Pontius}
\affiliation{Helmholtz-Zentrum Berlin f\"ur Materialien und Energie GmbH, Albert-Einstein-Stra\ss e 15, 12489 Berlin, Germany}
\author{Yuanhua Lin}
\affiliation{State Key Lab of New Ceramics and Fine Processing, School of Materials Science and Engineering, Tsinghua University, Beijing 100084, China}
\author{Cewen Nan}
\affiliation{State Key Lab of New Ceramics and Fine Processing, School of Materials Science and Engineering, Tsinghua University, Beijing 100084, China}
\author{Hiroki Wadati}
\affiliation{Department of Material Science, Graduate School of Science, University of Hyogo, Ako, Hyogo 678-1297, Japan}
\affiliation{Institute of Laser Engineering, Osaka University, Suita, Osaka 565-0871, Japan}

\begin{abstract}
We have observed photocarrier transport behaviors in BiFeO$_3$/La$_{1-x}$Sr$_x$MnO$_3$~(BFO/LSMO) heterostructures by using time-resolved synchrotron x-ray magnetic circular dichroism in reflectivity. The magnetization of LSMO layers was used as a probe of photo-induced carrier dynamics in the photovoltaic BFO layers. During the photo-induced demagnetization process, the decay time of LSMO~($x$=0.2) magnetization strongly depends on the ferroelectric polarization direction of the BFO layer. The variation of decay time should be attributed to the different sign of accumulated photocarriers at the BFO/LSMO interface induced by the photovoltaic effect of the BFO layer. The photocarriers can reach the BFO/LSMO interface and influence the magnetization distribution in the LSMO layers within the timescale of $\sim$100~ps. Our results provide a novel strategy to investigate carrier dynamics and mechanisms of optical control of magnetization in thin film heterostructures.

\end{abstract}

\maketitle

\section{Introduction}
Ferroelectric photovoltaic (FEPV) effect~\cite{han2022ferroelectric,paillard2016photovoltaics,fridkin1978anomalous,martin2016thin} has attracted a great amount of research attention due to its unique advantages of over-bandgap photovoltage and switchable photocurrent, compared to the conventional photovoltaic effect in p-n junction-based solar cells. On one hand, FEPV materials can generate photovoltage larger than their bandgaps, which is especially helpful for application of narrow-bandgap FEPV materials~\cite{yang2010above,seidel2011efficient}. On the other hand, ferroelectric~(FE) materials exhibit switchable electric polarization, which can couple with the direction of the photocurrent, adding much flexibility for manipulation of FEPV effects in photovoltaic devices~\cite{choi2009switchable,yang2009photovoltaic,zhou2020switchable,ji2010bulk}. 

Meanwhile, multiferroic heterostructures exhibiting both ferroelectricity and ferromagnetism~\cite{hu2016multiferroic,vaz2012electric,chen2022signatures,wu2021self} have been intensely investigated for a few decades, aiming at both interfacial magnetoelectric coupling mechanisms~\cite{jia2014mechanism,yao2015magnetoelectric} and potential applications such as memory devices~\cite{scott2007multiferroic,roy2012multiferroic}, sensors~\cite{vopson2015fundamentals}, etc. Generally, the manipulation of magnetism in multiferroic heterostructures can be realized by the strain, interfacial charge accumulation and interfacial exchange coupling accompanied with the switching of the electric polarization. Light excitation can act as an additional route to control the magnetism and transport properties in multiferroic heterostructures when the FE layer exhibits FEPV effect. It has been reported that generation and transport of photocarriers into the ferromagnetic(FM) layers can modify their magnetism and electric transport behaviors~\cite{zhao2019polarization,sung2014photo,zheng2018optically}. Nevertheless, the dynamic process of the carrier transport is rarely investigated in multiferroic heterostructures with FEPV effect.

BiFeO$_3$~(BFO)~\cite{yang2010above,choi2009switchable,seidel2011efficient,yang2009photovoltaic,zhou2020switchable,ji2010bulk,you2018enhancing} is one of the most intensely studied FEPV materials with outstanding FEPV properties. Here we select BFO/(La,Sr)MnO$_3$~(LSMO) multiferroic heterostructure as a model system to investigate the effects of photocarrier transport upon the magnetism of LSMO layer in time domain. Photocarrier transport dynamics is detected by element-specific time-resolved x-ray magnetic circular dichroism in reflectivity~(XMCDR). Finite penetration of XMCDR provides sensitivity to the depth profile of magnetization, and element specificity of XMCDR ensures that the magnetic signal comes from the LSMO layer. The photo-induced magnetic dynamics of LSMO layer strongly depend on the direction of the FE polarization of BFO layer, as well as the Sr concentration in LSMO layer, which can be explained by transient carrier accumulation at the multiferroic interface.

\section{Methods and Basic Sample Characterizations}

Epitaxial BFO/La$_{1-x}$Sr$_x$MnO$_3$~(LSMO$_x$, $x=$0.2 and 0.33) thin film heterostructures and SrTiO$_3$~(STO)/LSMO$_x$~($x=$0.2 and 0.33) reference samples were fabricated by pulsed laser deposition. LSMO$_x$ and BFO~(or STO) layers were grown in sequence on STO(001) single-crystalline substrates. The nominal thickness of the BFO, STO, LSMO$_{0.2}$ and LSMO$_{0.33}$ layers are 40, 40, 20 and 6 nm, respectively. The detailed growth parameters were reported elsewhere~\cite{zhao2019polarization}. 

\begin{figure}[t]
	\includegraphics[width=80 mm]{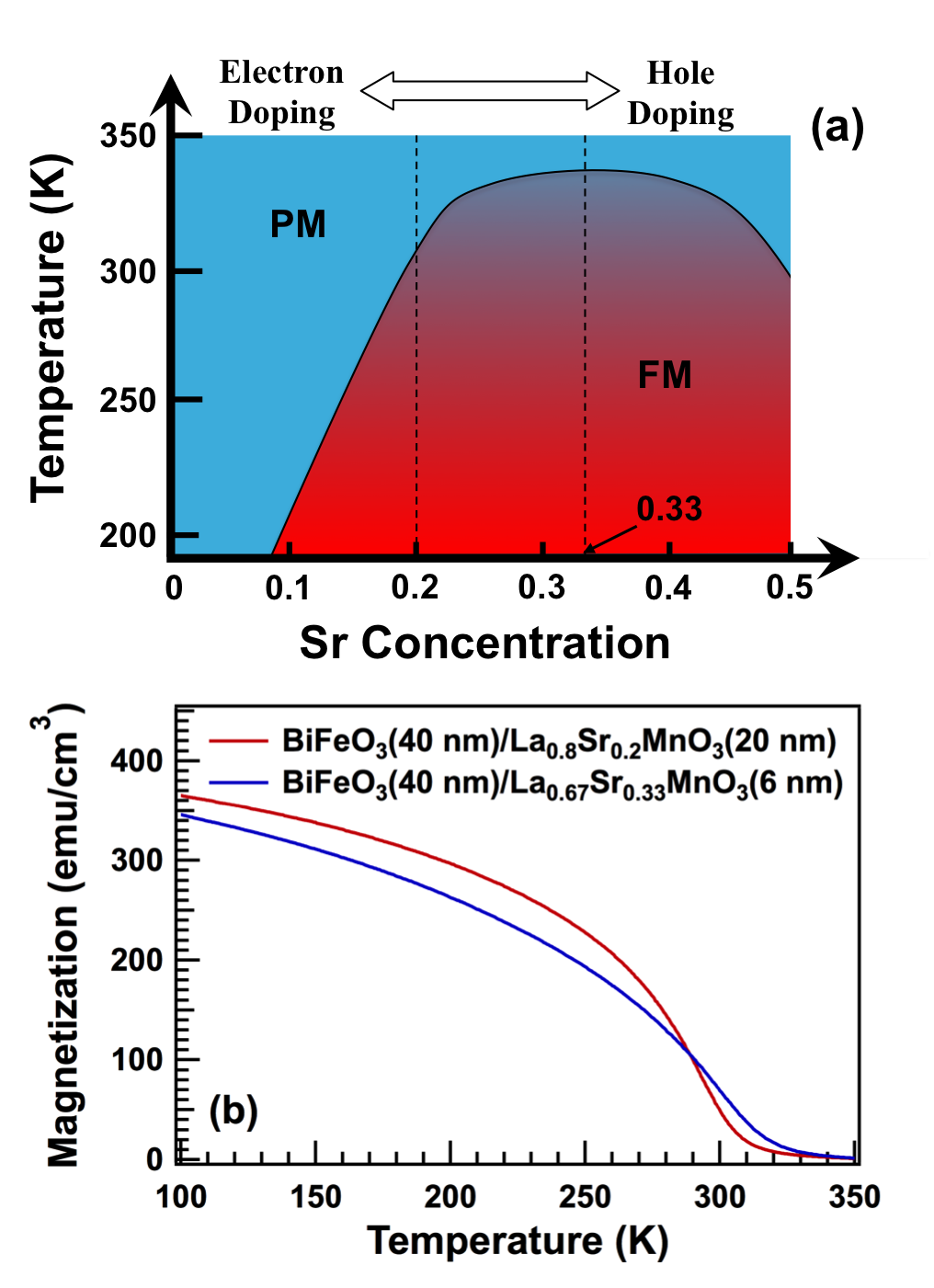}
	\caption{(a) Schematic phase diagram of LSMO. (b) $M-T$ curves of the BFO/LSMO multiferroic heterostructures.
		\label{fig1}
	}
\end{figure}

As schematically shown in Fig.~\ref{fig1}(a), the paramagnetic~(PM)-FM transition temperature~($T_C$) of LSMO depends on the Sr concentration~\cite{fujishiro1998charge,hemberger2002structural}. LSMO$_{0.2}$ has slightly lower $T_C$ than that of LSMO$_{0.33}$, as confirmed by the magnetization-temperature ($M-T$) curves in Fig.~\ref{fig1}(b). The magnetometry measurements were conducted by a superconducting quantum interference device (SQUID, Quantum Design).

\begin{figure}[t]
	\includegraphics[width=\columnwidth]{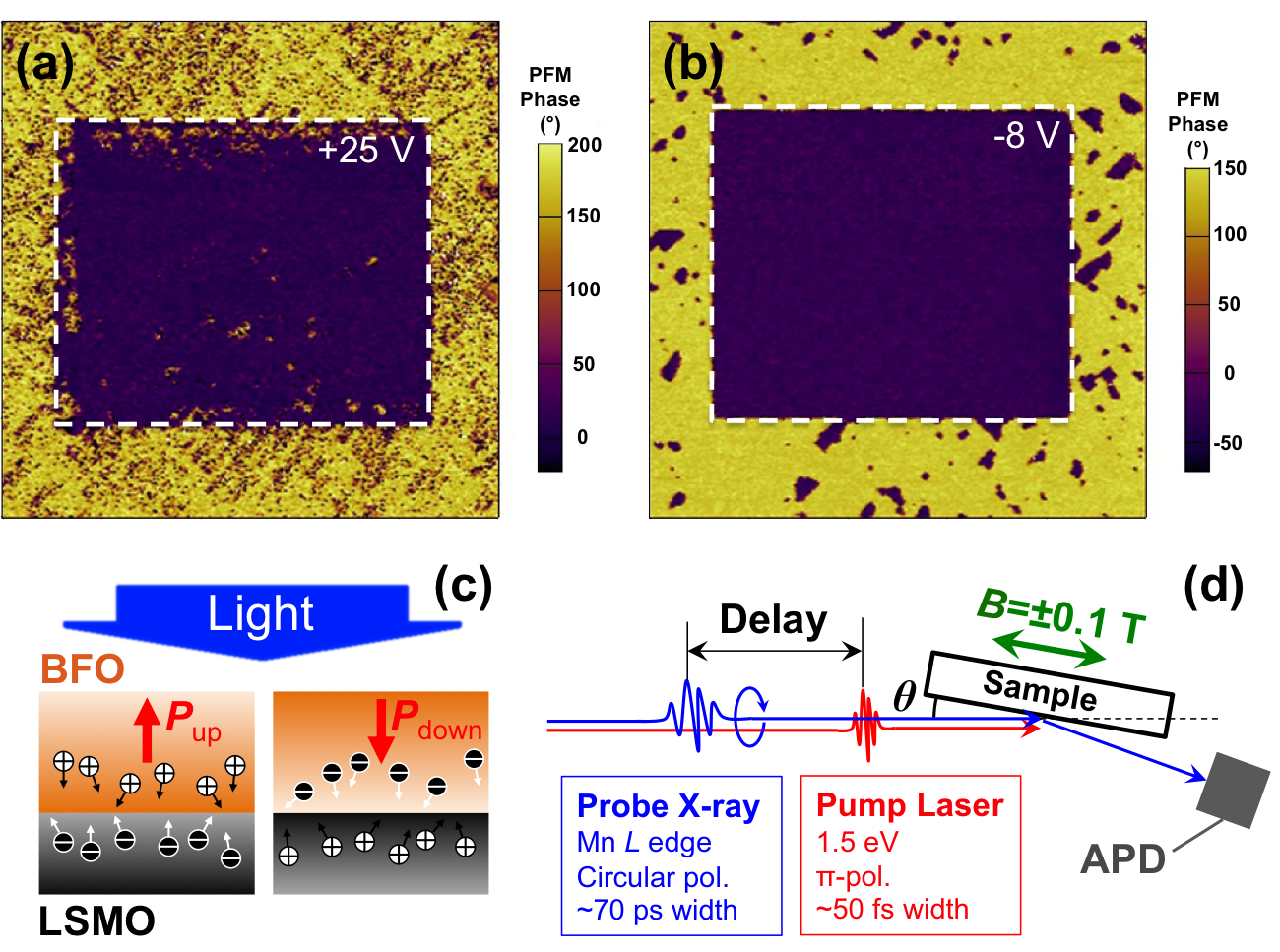}
	\caption{Out-of-plane PFM images of the BFO/LSMO multiferroic heterostructures with (a) ``up" and (b) ``down" FE polarization. The regions enclosed by dashed line were scanned by applying +25 V and -8 V tip bias respectively before the PFM measurements. (c) Schematic of the FEPV-effect-induced carrier transport in BFO/LSMO heterostructures. The deeper and lighter colors indicate the accumulation of positive and negative charge respectively. (d) Setup of the time-resolved XMCDR experiment.
		\label{fig2}
	}
\end{figure}

The pristine out-of-plane polarization of the BFO layers is pointing out of the film surface (defined as polarization ``up", $P_{up}$), as confirmed by the piezoelectric force microscopy~(PFM) results shown in Fig.~\ref{fig2}(a). The out-of-plane polarization of BFO can be switched to ``down" ($P_{down}$, pointing into the film surface, Fig.~\ref{fig2}(b)) by applying DC voltage in water~\cite{zhao2019polarization,tian2018water}. The PFM experiments were performed in ambient conditions at room temperature with an atomic force microscopy system~(Infinity, Asylum Research). The STO-capped samples act as zero-polarization references to compare with the BFO/LSMO samples.

Optical excitation above the bandgap of BFO ($\sim$2.8 eV~\cite{sando2018revisiting}) is expected to generate free photocarriers. The internal electric field of FE polarization can separate the photocarriers and drive electrons and holes to opposite directions. The photocarriers transported to the BFO/LSMO interface can break the local electric balance, leading to the modulation of carrier density in the LSMO layer, as schematically shown in Fig.~\ref{fig2}(c). 

Here we design pump-probe experiments to investigate the dynamic modulation of LSMO magnetism by the pump laser. The time-resolved XMCDR measurements were carried out at beamline UE56/1-ZPM (FEMTOSPEX) of BESSY~II by using the setup shown in Fig.~\ref{fig2}(d). \mbox{X-ray} with fixed circular polarization at the Mn $L$ edge was used and an in-plane magnetic field of $\pm$0.1~T was switched to observe the magnetic contrast in reflectivity. 
The reflectivity was detected with an avalanche photodiode~(APD) and boxcar integrated. A Ti:sapphire laser (frequency doubled, $\lambda$=400~nm, $h\nu\sim3$~eV, $\pi$-polarization, 3~kHz, pulse width $\sim$50~fs) was employed as the pump source. For the measurement of LSMO$_{0.2}$ sample, the full width at half maximum~(FWHM) spot sizes (horizontal$\times$vertical) of the pump laser and the probe \mbox{x-ray} were around 0.19$\times$0.14~mm$^2$ and 0.12$\times$0.12~mm$^2$, respectively. For the measurement of LSMO$_{0.33}$ sample, the FWHM spot sizes (horizontal$\times$vertical) of the pump laser and the probe \mbox{x-ray} were around 0.22$\times$0.28~mm$^2$ and 0.11$\times$0.12~mm$^2$, respectively. The time resolution of the measurements was limited to $\sim$70~ps by the pulse width of the probe \mbox{x-ray}. The pumped and unpumped signals were collected alternatively by recording the contributions from the pumped and unpumped bunches. The samples were cooled down to 200~K by a liquid N$_2$ flow cryostat. All the laser fluences mentioned below are calibrated fluences which are absorbed by the samples.

According to the reported optical properties of BFO~\cite{xu2009optical,vzelezny2010optical,liu2013strain} and STO~\cite{du2003optical,thomas2000optical,gao2003band}, 40~nm of BFO can absorb most ($\sim$95\%) of the 400~nm pump laser, while STO, whose bandgap is $\sim$3.7~eV, can only absorb $<$10\% of the 400~nm pump laser, when taking the incident angle of $\theta$ =15\textsuperscript{o} into consideration. Most of the pump fluence was absorbed by BFO layer in BFO/LSMO samples, while for STO/LSMO reference samples, the LSMO is nearly directly pumped by the 400~nm laser. Thus, the magnetic dynamics of BFO/LSMO samples observed in our setup mainly reflects the effects induced by optical pumping of BFO layer. 
The penetration depth of x-ray at the Mn $L$ edges is comparable for BFO and STO (estimated by CXRO~\cite{CXRO}), thus the probing depth of XMCDR is similar for all the samples.

The calculation of XMCDR was conducted by the ReMagX software~\cite{ReMagX}. The non-resonant optical constants~(real and imaginary parts of the refractive index, $\delta$ and $\beta$) of BFO, STO and LSMO were obtained from the optical database of Henke~\cite{henke1993x,CXRO}. For LSMO layers, the imaginary part of the refractive index $\beta$ as well as its magnetic dichroism $\beta_M$ at the Mn $L$ edge were extracted from Ref.~\cite{aruta2009orbital}~(ignoring the Sr-concentration dependence of the spectral shape) and scaled to fit into the optical constant data from the Henke's database. Consequently, Kramers-Kronig transformation was conducted on the imaginary parts to obtain the real parts of the optical constants (including the magnetic real part, $\delta_M$). For all the calculations, the roughness of all the layers were set as zero, and the angular and energy resolution was set as 5~mrad and 1~eV, respectively.

\section{Results and Discussions}

\begin{figure}[t]
	\includegraphics[width=\columnwidth]{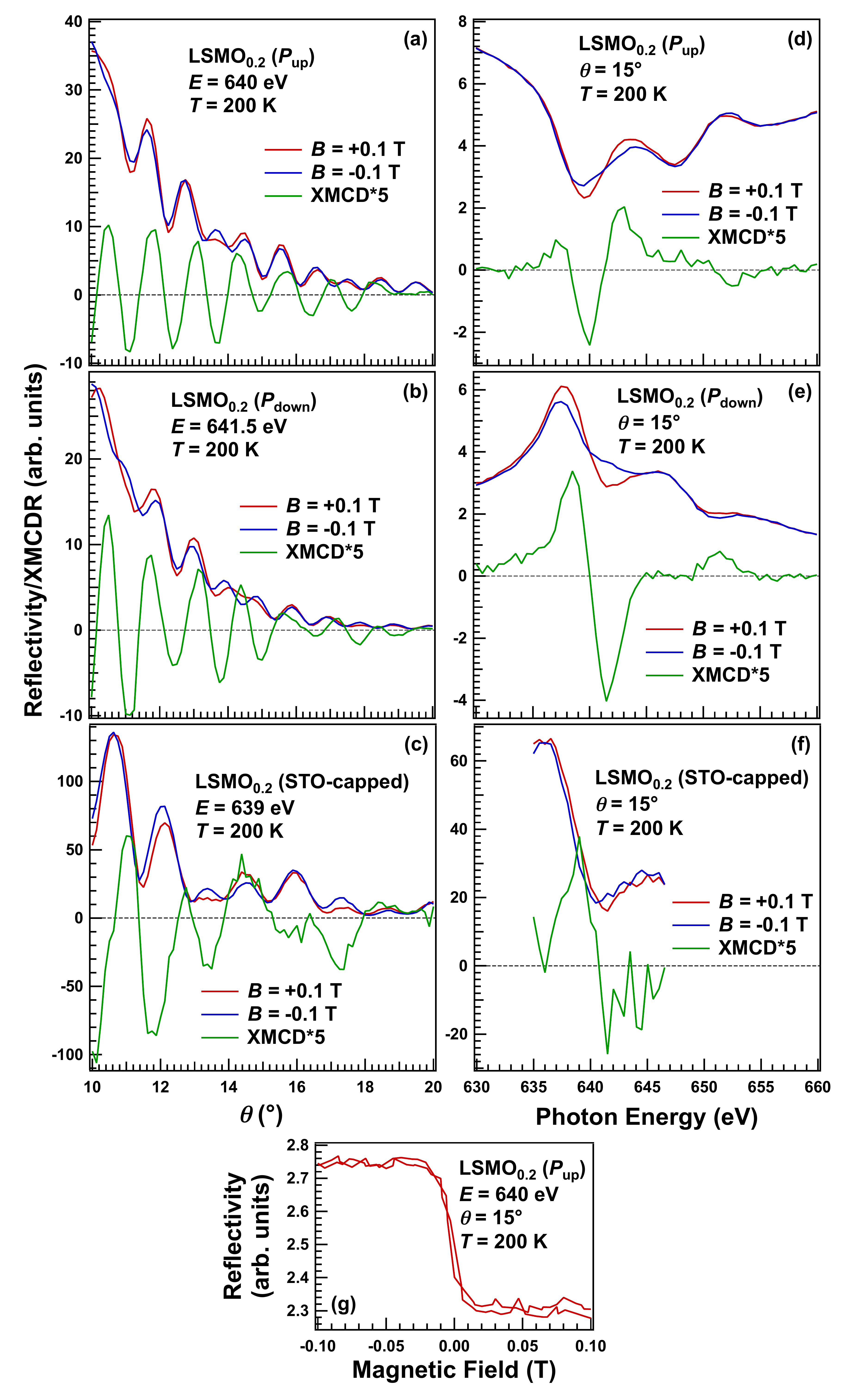}
	\caption{Static soft x-ray reflectivity and XMCDR of LSMO$_{0.2}$ sample at the Mn $L$ edges. (a) $\theta-2\theta$ scan, $P_{up}$; (b) $\theta-2\theta$ scan, $P_{down}$; (c) $\theta-2\theta$ scan, STO-capped; (d) Energy scan, $P_{up}$; (e) Energy scan, $P_{down}$; (f) Energy scan, STO-capped; (g) Element specific magnetic hysteresis loop. 
		\label{fig3}
	}
\end{figure}

XMCDR at the Mn $L$ edges contains information about the Mn magnetization near the BFO/LSMO interface. Fig.~\ref{fig3} shows the static reflectivity and XMCDR of the LSMO$_{0.2}$ samples with $P_{up}$ and $P_{down}$, as well as the STO-capped sample. Oscillations of the specular reflectivity ~(Fig.~\ref{fig3}(a-c)) indicate perfect surface and interface quality of the heterostructures. XMCDR also exhibits oscillations and an incident angle of $\theta$=15\textsuperscript{o} was chosen for the following time-resolved measurements to obtain the best magnetic contrast. Energy scans at the Mn $L$ edges show significant circular dichroism of the reflectivity signal~(Fig.~\ref{fig3}(d-f)). The time-resolved measurements were conducted at optimized photon energies with large XMCDR for each sample. The magnetic hysteresis of the reflectivity signal shown in Fig.~\ref{fig3}(g) confirms the FM nature of the LSMO layer. The LSMO$_{0.33}$ samples show similar properties as that of the LSMO$_{0.2}$ samples.

\begin{figure}[t]
	\includegraphics[width=70 mm]{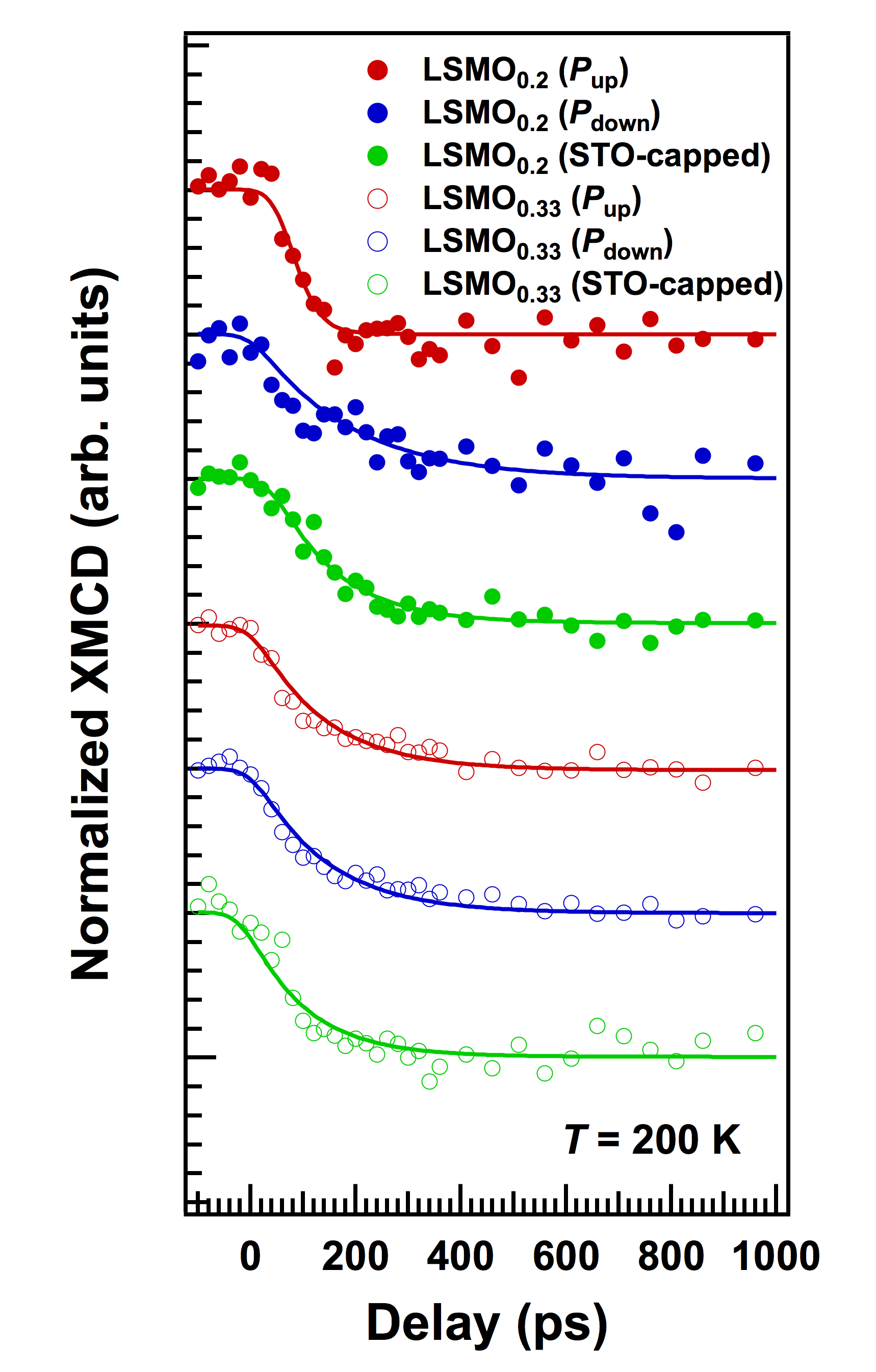}
	\caption{Magnetic dynamics of the heterostructure samples. The solid lines shows the fitting results by using the function (1). Pump effects of all the samples are normalized to 1 by scaling the $I_0$=$I_1$=1 and the data are vertically shifted. The measurements of LSMO$_{0.2}$ and LSMO$_{0.33}$ samples were conducted with pump fluence of 20 and 11.65~mJ/cm$^2$, respectively. The incident angle was kept at $\theta$ =15\textsuperscript{o} and the photon energy varied for different samples and was chosen to maximize the XMCDR signal.
		\label{fig4}
	}
\end{figure}

Light excitation can cause demagnetization of FM materials. According to the schematic magnetic phase diagram of LSMO shown in Fig.~\ref{fig1}(a), the temperature of PM-FM transition of LSMO strongly depends on the Sr doping concentration at $x\sim0.2$ while exhibits weak dependence on the Sr concentration at $x\sim0.33$. Thus it is expected that the magnetic dynamics of LSMO$_{0.2}$ samples will be significantly influenced by the out-of-plane polarization direction of the BFO layer, whereas that of the LSMO$_{0.33}$ samples will be barely influenced by the FEPV effect of the BFO layer. These expectations were confirmed by the time-resolved XMCDR measurements depicted in Fig.~\ref{fig4}. The delay scans are fitted by the function
$$I(t)=I_0-I_1 \mathrm{exp} (-t/\tau_{decay})H(t)\;\; (1)$$
convolved with a 70-ps-wide Gaussian time-resolution function. $H(t)$ is the Heaviside step function and the parameter $\tau_{decay}$ is used for evaluation of the demagnetization timescale.

For the LSMO$_{0.2}$ samples, ``up" polarization of the BFO obviously induced a faster demagnetization of the LSMO layer, while the ``down" polarization did the opposite effect. The $\tau_{decay}$ of $P_{up}$, $P_{down}$ and STO-capped samples are determined as 39.9, 122.3 and 82.3~ps, respectively. While for the LSMO$_{0.33}$ samples, the $\tau_{decay}$ of $P_{up}$, $P_{down}$ and STO-capped samples are determined as 70.3, 70.0 and 69.6~ps, respectively, indicating the negligible role of BFO polarization on the magnetic dynamics of Mn magnetization in LSMO$_{0.33}$ samples. STO capping layers barely absorb the 400~nm pump laser, hence the magnetic dynamics of the STO-capped samples can be regarded as a LSMO single layer reference. The distinct magnetic dynamics of LSMO$_{0.2}$ samples with different BFO polarization should be attributed to the fact that the PM-FM phase transition temperature changes steeply with the Sr concentration for LSMO$_{0.2}$~(left dashed line in Fig.~\ref{fig1}(a)), and electron/hole doping does opposite effect on the magnetization. While the PM-FM phase transition temperature of LSMO$_{0.33}$ is near the maximum~(right dashed line in Fig.~\ref{fig1}(a)) when changing the Sr concentration. Either electron or hole doping will induce similar decrease of magnetization in LSMO$_{0.33}$ samples.

\begin{figure}[t]
	\includegraphics[width=\columnwidth]{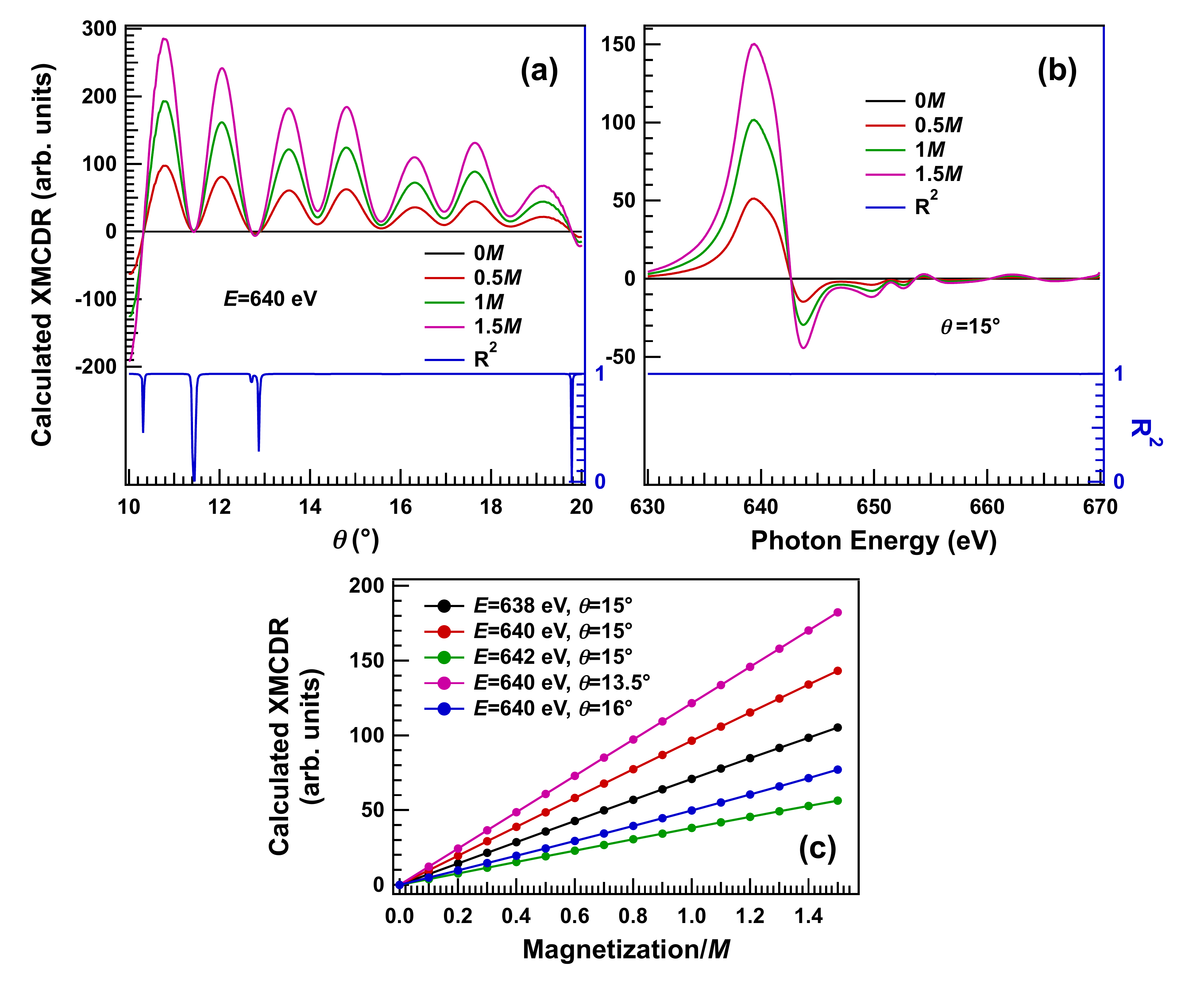}
	\caption{Calculated (a) angle and (b) energy dependence of XMCDR with different size of magnetization in LSMO$_{0.2}$ layer of the BFO(40~nm)/LSMO$_{0.2}$(20~nm)/STO heterostructure. The left axes show the calculated XMCDR and the right axes show the R-square value obtained from linear fitting of calculated XMCDR against the magnetization. (c) Dependence of calculated XMCDR against the LSMO magnetization~(in the unit of static magnetization of LSMO$_{0.2}$, $M$) at selected energy and reflection angle. 
		\label{fig5}
	}
\end{figure}

Due to the difference of the angle and energy profile of the reflectivity, the absolute value of the pump effect can vary. Thus in Fig.~\ref{fig4} we normalized the maximum pump effect to the same value for comparison. 
We have confirmed by theoretical calculation that the dynamic behaviors of XMCDR should hardly depend on the reflection angle or photon energy, as depicted in Fig.~\ref{fig5}. Assuming the static magnetization of LSMO$_{0.2}$ as $M$, by varying the size of magnetization from 0$M$ to 1.5$M$ (by scaling the energy dependent $\beta_M$) in the BFO(40~nm)/LSMO$_{0.2}$(20~nm) sample, the angular and energy dependence of the XMCDR exhibit similar shape (Fig.~\ref{fig5}(a,b), left axes) and different size. Fig.~\ref{fig5}(c) shows the roughly linear dependence of calculated XMCDR upon the size of magnetization at various energy and reflection angle. By conducting linear fit of calculated XMCDR against the magnetization at different energy and reflection angle in Fig.~\ref{fig5}(a,b), the R-square values of the fitting is plotted on the right axes of Fig.~\ref{fig5}(a,b). It could be noticed that the linear relationship between XMCDR and the magnetization only breakdown at reflection angles when the XMCDR is close to zero. Thus, when choosing the energy and reflection angle at a local maximum of the XMCDR, such as cases shown in Fig.~\ref{fig3}, the linear relationship between XMCDR and $M$ is well preserved. Consequently, the transient XMCDR should only be proportional to the magnetization of LSMO and independent on the reflection angle and photon energy. Note that the calculated angle/energy dependence of XMCDR differs from the real experimental data, because the angle/energy dependence of resonant reflectivity and XMCDR is very sensitive to the thickness/roughness of distinct layers.

\begin{figure}[t]
	\includegraphics[width=\columnwidth]{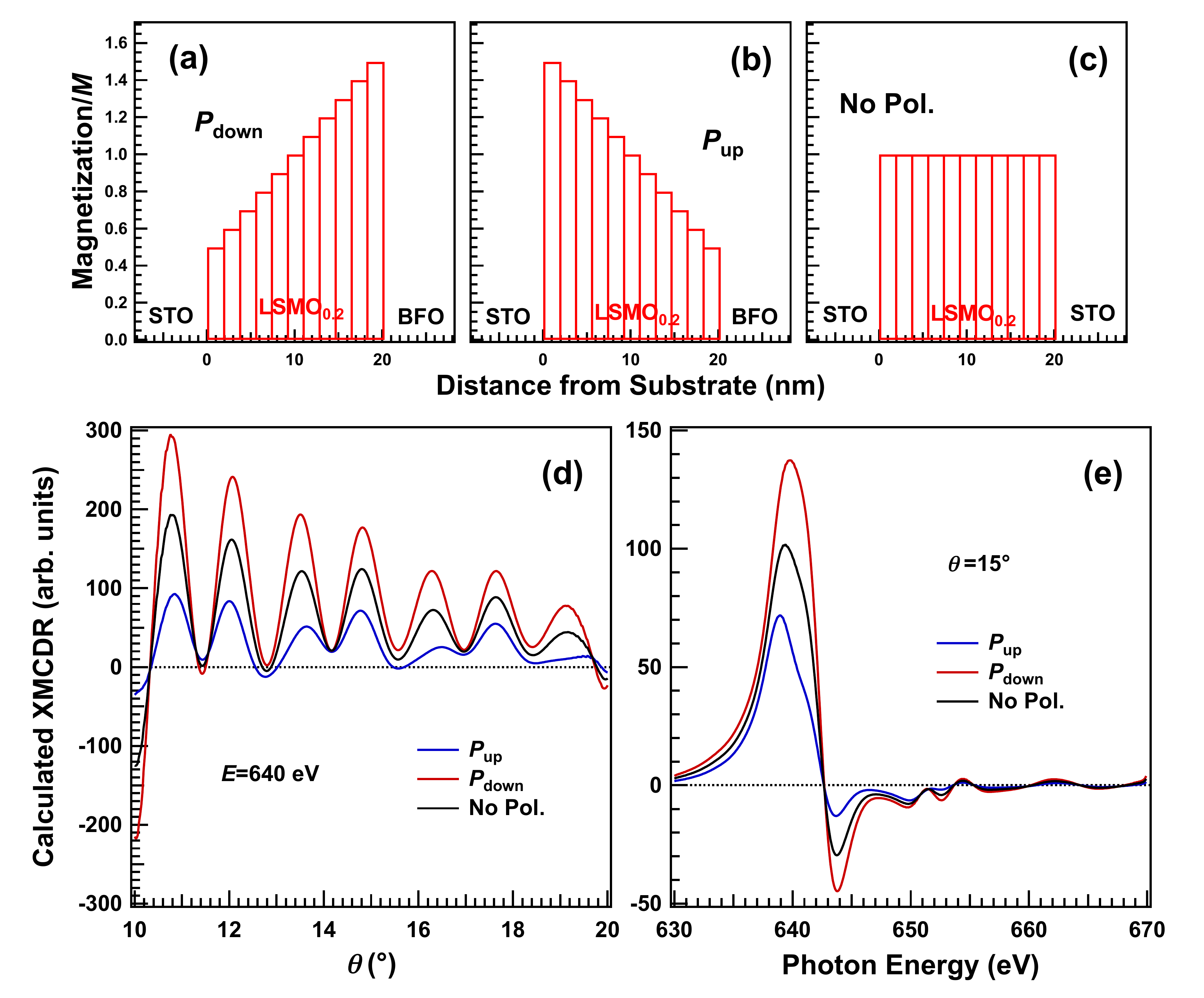}
	\caption{(a,b,c) Supposed transient magnetization profiles of the LSMO$_{0.2}$ layer caused by FEPV-induced charge redistribution. Calculated (d) angle and (e) energy dependence of the transient XMCDR using the supposed magnetic profiles. 
		\label{fig6}
	}
\end{figure}

According to the modification of magnetic dynamics induced by the BFO polarization, the detailed photocarrier transport behavior can be clarified. The photocarrier induced by 400~nm light illumination can be transported to the BFO/LSMO interface by FEPV effect. For the $P_{up}$ samples, the positive charge will move towards the interface, while for the $P_{down}$ samples, the negative charge will move towards the interface. There are two possible regimes of the carrier transport at the BFO/LSMO interface. The photocarriers can either move across the interface and be injected into the LSMO layers, or accumulate at the interface, depending on the interfacial potential barrier. 

In case of the injection regime, due to the insulating nature of the STO substrate, $P_{up}$ will induce hole doping in the LSMO$_{0.2}$ layer, resulting in increase of magnetization, while $P_{down}$ injects electrons into the LSMO$_{0.2}$ layer, leading to decrease of its magnetization. This contradicts with the experimental observations in Fig.~\ref{fig5}, where $P_{up}$ accelerates the demagnetization and $P_{down}$ slows down the demagnetization, with respect to the STO-capped sample.

Thus, the accumulation regime should be considered to explain the magnetic dynamics shown in Fig.~\ref{fig4}. When the photocarriers in BFO layer accumulate at the BFO/LSMO interface, they can attract charge with the opposite sign in LSMO layer to the other side of the interface, as shown in Fig.~\ref{fig2}(c). Since the STO substrate is insulating and there is no external source of charge compensation. Conservation of charge induces charge redistribution in the LSMO layer. In the $P_{up}$ samples, negative charge~(electron doping) migrates to the BFO/LSMO interface and positive charge~(hole doping) remains at the LSMO/STO interface, resulting in formation of a transient charge gradient. The sign of charge redistribution is opposite for the $P_{down}$ samples. The depth profile of LSMO magnetization should follow the charge gradient. To simulate such thickness dependence of electron/hole doping, here we use simplified magnetic depth profiles~(Fig.~\ref{fig6}(a,b,c)) and theoretically calculated the corresponding  angle/energy dependence of XMCDR~(Fig.~\ref{fig6}(d,e)) to mimic the transient behaviors of XMCDR. It could be clearly observed that although the total magnetization of the LSMO$_{0.2}$ layer remains the same for the simulated ``$P_{up}$", ``$P_{down}$" and ``no polarization" cases, larger magnetization near the BFO/LSMO interface leads to significantly larger XMCDR signal. This is because the finite penetration depth of the Mn $L$ edge soft \mbox{x-ray} makes the magnetic moments at deeper position contribute less to the XMCDR signal. The accumulation regime well agrees with the observed magnetic dynamic behaviors in Fig.~\ref{fig4}.

\section{Conclusions}

In summary, we have observed dynamics of photocarriers by monitoring the magnetic dynamics of adjacent FM layers. Different out-of-plane FE polarization can drive the photocarriers with different signs to accumulate at the FE/FM interface. To compensate the transient charge accumulation in the FE layer, the free carriers with opposite charge sign in the FM layer migrate to the FE/FM interface, inducing a transient charge/magnetization redistribution in the FM layer. The time scale of these processes is around 100~ps. The FEPV-effect induced magnetization change superposes with the photo-induce demagnetization, resulting in the magnetic dynamic behaviors of the FE/FM heterostructures. Our results clarifies the photo-induced carrier transport behaviors at FE/FM interface, which should be useful for development of novel light-manipulated magnetic devices and related applications.

\section{Acknowledgements}
This work was supported by Natural Science Foundation of China (Grant No. 52002370), JSPS KAKENHI Grant No. 17F17327 and Basic Research Funding of IHEP (Grant No. Y9515560U1). We thank HZB for the allocation of synchrotron radiation beamtime (Proposals 202-09723ST/R, 201-09271ST, 191-07992ST and 192-08474ST/R). We acknowledge the helpful discussion with K. Takubo and K. Yamamoto from University of Tokyo, as well as experimental supports provided by Karsten Holldack and Rolf Mitzner from BESSY II, Prof. J. X. Zhang from Beijing Normal University, Prof. J. Ma and Dr. M. F. Chen from Tsinghua University.

\bibliography{ref}

\end{document}